\documentclass[11pt,a4paper]{article}

\usepackage[margin=1in]{geometry} 
\usepackage{amsmath}
\usepackage{amssymb}
\usepackage{amsfonts}
\usepackage{graphicx}
\usepackage{bm}
\usepackage{mathtools}
\usepackage{xcolor}
\usepackage{hyperref}

\newcommand{\BH}{\mathrm{BH}}
\newcommand{\DM}{\mathrm{DM}}
\newcommand{\ISCO}{\mathrm{ISCO}}

\newcommand{\Hay}{\mathrm{Hay}}
\newcommand{\Kerr}{\mathrm{Kerr}}
\newcommand{\Msun}{M_{\odot}}

\begin{document}
	
	\title{Breaking the Degeneracy: Spectral Hardening of Accretion Disks around Rotating Hayward Black Holes in Dark Matter Halos}
	
	\author{
		Sandip Dutta \\[1ex]
		\small Department of Applied Mathematics, Dinabandhu Andrews Institute \\
		\small of Technology and Management, Kolkata, West Bengal, India \\[0.5ex]
		\small \texttt{duttasandip.mathematics@gmail.com}
	}

	\maketitle
	
	\begin{abstract}
		We investigate the thermodynamic and observable signatures of thin accretion disks surrounding rotating, regular Hayward black holes embedded within a macroscopic dark matter (DM) envelope. The spacetime is rigorously modelled as a three-region composite: an inner Hayward core regularised by a de~Sitter limit, an intermediate DM shell modelled as a pressureless dust envelope governed by an exponential sphere density profile, and an asymptotically flat outer vacuum. By matching these regions via the mass profile, we compute the explicit modifications to the Innermost Stable Circular Orbit (ISCO), radiative efficiency, local thermal flux, and multi-colour blackbody spectral luminosity. A systematic comparison across four distinct configurations (Kerr vacuum, Hayward vacuum, Kerr\,+\,DM, and Hayward\,+\,DM) reveals a strict hierarchical compression of the ISCO. We demonstrate that the quantum-inspired core regularity and the macroscopic DM halo exert an additive enhancement on the radiative efficiency, reaching $\sim 14.4\%$ for highly spinning black holes ($j=0.8$). However, this additive behaviour introduces a profound macroscopic degeneracy between purely geometric (Hayward vacuum) and purely astrophysical (Kerr\,+\,DM) configurations at low-to-intermediate frequencies. We establish that this degeneracy is structurally broken only in the extreme Wien tail of the multi-wavelength spectrum, providing a critical diagnostic footprint for future high-frequency spectropolarimetry to distinguish non-singular black holes from classical dark matter environments.
	\end{abstract}
	
	\vspace{1em}

\section{Introduction}
\label{sec:intro}

General relativity predicts the inevitability of spacetime singularities inside black holes (BHs) under reasonable physical conditions. However, it is widely expected that quantum-gravity effects will regularise these classical singularities, replacing them with non-singular, de~Sitter-like cores at the Planck scale. Regular (non-singular) BH solutions provide an effective macroscopic description of this expectation within classical general relativity. Since Bardeen's construction of the first regular solution \cite{Bardeen1968}, the theoretical investigation of non-singular geometries has expanded significantly, driven by couplings to non-linear electrodynamics (NLE) and modified gravity theories \cite{Dymnikova1992,AyonBeato1998,Bronnikov2000,Malafarina2022}. 
Among these models, the Hayward regular black hole \cite{Hayward2006} occupies a central position in the literature. Its mass function, $m(r)=Mr^{3}/(r^{3}+r_{0}^{3})$, is the simplest construct that simultaneously guarantees asymptotic flatness, a regular origin, and a singularity-free de~Sitter interior. The rotating extension of the Hayward metric, derived via the Azreg-A\"{i}nou non-complexification algorithm \cite{AzregAinou2014}, preserves the NLE matter content of the static solution and has been subjected to detailed scrutiny regarding its horizon structure and thermodynamics \cite{Bambi2013,Toshmatov2014}. 

A primary method for testing the astrophysical viability of these alternative metrics is through the electromagnetic signatures of their accretion disks. Recently, Boshkayev et al. (2024) \cite{Boshkayev2024} computed the full Novikov-Thorne thin accretion disk luminosity around rotating Hayward BHs in a vacuum. They found significant deviations from the standard Kerr spectrum, particularly at intermediate values of the regularity parameter, identifying the inclusion of dark matter (DM) as a critical next step for realistic astrophysical modeling. Independently, the profound influence of DM halos on accretion disk spectra has been extensively studied for static, singular BHs. These investigations have modelled the DM envelope under various thermodynamic assumptions, including isotropic pressure \cite{Boshkayev2020}, anisotropic pressure \cite{Kurmanov2022}, and tangential-only pressure (the Einstein-cluster model) \cite{Boshkayev2022}. 

Despite the extensive separate investigations into regular black holes and dark matter environments, the combination of a regular BH core embedded within a macroscopic DM envelope has not been explored in the literature. Previous studies investigating DM-immersed accretion disks have universally assumed a singular (Schwarzschild or Kerr) spacetime at the core. Consequently, the interplay between the short-range quantum-inspired regularity parameter and the long-range macroscopic DM density profile remains unknown.
The primary motivation for this research is to ascertain whether the phenomenological effects of a de Sitter-like regular core and an external DM halo produce degenerate, competing, or additive signatures on the observable accretion disk luminosity. Because both alternative gravity effects and dark matter environments can mimic the spin of a standard Kerr black hole by shifting the Innermost Stable Circular Orbit (ISCO), it is imperative to determine if multi-wavelength spectral luminosity can break this parameter degeneracy. 
To address this gap, we construct a unified theoretical framework comprising a three-region spacetime: (i) an inner rotating Hayward BH core ensuring a regular interior; (ii) an intermediate DM shell governed by a pressureless dust envelope using an exponential sphere density profile \cite{Sofue2013}; and (iii) an outer Kerr vacuum. By matching these regions via the mass profile, we derive a modified Boyer-Lindquist metric with a running mass. We apply the standard Novikov-Thorne-Page (NTP) accretion disk formalism \cite{Novikov1973,Page1974} to extract the orbital mechanics, radiative flux, effective temperature, and multi-colour blackbody spectral luminosity.

To systematically isolate the effects of regularity and DM, we conduct a unified four-case comparison:
\begin{enumerate}
	\item \textbf{Kerr vacuum} ($r_{0}^{*}=0$, $\rho_{0}=0$): standard NTP baseline.
	\item \textbf{Rotating Hayward vacuum} ($r_{0}^{*}>0$, $\rho_{0}=0$): reproducing Ref.~\cite{Boshkayev2024}.
	\item \textbf{Kerr + DM} ($r_{0}^{*}=0$, $\rho_{0}>0$): extending Refs.~\cite{Kurmanov2022,Boshkayev2022} to spinning BHs.
	\item \textbf{Rotating Hayward + DM} ($r_{0}^{*}>0$, $\rho_{0}>0$): the novel configuration proposed in this work.
\end{enumerate}

The paper is organised as follows. Section~\ref{sec:hayward} establishes the algebraic and geometrical structure of the rotating Hayward metric. Section~\ref{sec:spacetime} constructs the three-region spacetime architecture and the macroscopic dark matter envelope. Section~\ref{sec:TOV} formulates the metric construction and the pressureless dust halo. Section~\ref{sec:disk_dynamics} derives the orbital dynamics of test particles and the subsequent accretion disk thermodynamics. Section~\ref{sec:methods} outlines the numerical methodology, parameter space, and validation benchmarks. Section~\ref{sec:results} provides a detailed diagrammatic interpretation of the observational signatures. Finally, Section~\ref{sec:conclusions} summarises our conclusions. Geometric units $G=c=1$ are used throughout.

\section{Algebraic and Geometrical Structure of the Rotating Hayward Spacetime}
\label{sec:hayward}

The Hayward regular black hole \cite{Hayward2006} is an exact topological solution obtained by coupling general relativity to a non-linear electrodynamics (NLE) source, where the Lagrangian is precisely formulated to guarantee global regularity. The defining geometrical characteristic of this spacetime is encapsulated by its mass function,
\begin{equation}
	\label{eq:mhay}
	m(r) = M\,\frac{r^{3}}{r^{3} + r_{0}^{3}},
\end{equation}
where $M$ represents the ADM mass evaluated at spatial infinity, and $r_{0}$ is a fundamental characteristic length scale parameterizing the deviation from the classical singular Schwarzschild geometry. In dimensionless terms, this is expressed as $r_{0}^{*}\equiv r_{0}/M$. The profile in Eq.~(\ref{eq:mhay}) is a specific realization ($p=q=3$) of a broader family of NLE exact solutions \cite{AyonBeato1998,Bronnikov2000}:
\begin{equation}
	\label{eq:mgeneral}
	m(r) = M\!\left[1 + \left(\frac{r_{0}}{r}\right)^{q}\right]^{-p/q}.
\end{equation}
The physical integrity of this formulation rests on its asymptotic and boundary behaviors. As $r\to\infty$, the mass function smoothly recovers the asymptotic flatness of the classical limit, $m(r)\to M$. Conversely, in the deep interior limit as $r\to0$, the core regularizes. The mass scales as $m(r)\approx Mr^{3}/r_{0}^{3}$, ensuring the temporal metric component behaves as:
\begin{equation}
	\label{eq:core}
	g_{tt} = 1 - \frac{2m(r)}{r} \xrightarrow{r\to0} 1 - \frac{2M}{r_{0}^{3}}\,r^{2}.
\end{equation}
This reveals a locally de~Sitter geometry governed by an effective cosmological constant $\Lambda_{\rm eff}=6M/r_{0}^{3}$. This fundamental replacement of the classical central singularity with a de~Sitter core ensures that curvature invariants, notably the Kretschmann scalar $\mathcal{K}=R_{\mu\nu\rho\sigma}R^{\mu\nu\rho\sigma}$, remain strictly finite globally.

To model astrophysically relevant accretion processes, this static geometry must be extended to encompass angular momentum. The rotating extension, characterized by the spin parameter $a=jM$ (where $|j|\le 1$), is rigorously derived via the Azreg-A\"{i}nou non-complexification algorithm \cite{AzregAinou2014}. In standard Boyer--Lindquist coordinates $(t,r,\theta,\phi)$, the axisymmetric line element is formulated as:
\begin{equation}
	\label{eq:metric}
	ds^{2} = -\frac{\Delta_{r}}{\Sigma}\!\left(dt - a\sin^{2}\!\theta\,d\phi\right)^{\!2}
	+ \frac{\Sigma}{\Delta_{r}}\,dr^{2}
	+ \Sigma\,d\theta^{2}
	+ \frac{\sin^{2}\!\theta}{\Sigma}\!\left[(r^{2}+a^{2})\,d\phi - a\,dt\right]^{\!2},
\end{equation}
where the structurally modified horizon function $\Delta_{r}$ and the standard kinetic term $\Sigma$ are defined as:
\begin{align}
	\label{eq:Delta}
	\Delta_{r} &= r^{2} - 2m(r)\,r + a^{2},\\[3pt]
	\label{eq:Sigma}
	\Sigma     &= r^{2} + a^{2}\cos^{2}\!\theta.
\end{align}
By restricting the domain to the equatorial plane ($\theta=\pi/2$, $\Sigma=r^{2}$)—which serves as the primary locus for thin accretion disk dynamics—the non-vanishing covariant metric tensor components simplify to:
\begin{align}
	g_{tt}         &= -\!\left(1 - \frac{2m(r)}{r}\right), \label{eq:gtt}\\[4pt]
	g_{rr}         &= \frac{r^{2}}{\Delta_{r}},             \label{eq:grr}\\[4pt]
	g_{\phi\phi}   &= r^{2} + a^{2} + \frac{2m(r)\,a^{2}}{r}, \label{eq:gpp}\\[4pt]
	g_{t\phi}      &= -\frac{2m(r)\,a}{r},                  \label{eq:gtp}
\end{align}
yielding the standard equatorial three-volume element, $\sqrt{-g}=r$.

The causal structure and the existence of event horizons within this spacetime are governed by the roots of the modified horizon function, $\Delta_{r}=0$. Expanding this condition yields the quintic algebraic equation:
\begin{equation}
	\label{eq:horizon_poly}
	r^{5} - 2Mr^{4} + a^{2}r^{3} + r_{0}^{3}r^{2} + a^{2}r_{0}^{3} = 0.
\end{equation}
In the static limit ($a=0$), this structural equation reduces to a cubic polynomial:
\begin{equation}
	\label{eq:horizon_static}
	r^{3} - 2Mr^{2} + r_{0}^{3} = 0.
\end{equation}
This cubic admits two distinct positive real roots—the inner Cauchy horizon $r_{-}$ and the outer event horizon $r_{+}$—provided that the regularity parameter remains below a critical threshold. The degenerate case, where the horizons merge ($r_{-}=r_{+}$), defines the extremal static configuration exactly at:
\begin{equation}
	\label{eq:extremal}
	r_{0,\rm ext} = \frac{4M}{3\sqrt{3}} \approx 0.770\,M.
\end{equation}
For the general rotating case ($a\ne0$), the extremality condition tracing the bifurcation of the roots must be evaluated numerically (see Figure~\ref{fig:horizons}). For coordinate combinations exceeding this boundary, the spacetime transitions into a horizonless, globally regular compact object.

\section{Three-Region Spacetime Architecture and Dark Matter Envelope}
\label{sec:spacetime}

To physically model an astrophysical environment where a regular black hole is embedded within a galactic dark matter halo, we construct a composite spacetime divided into three concentric regions. The innermost region ($r < r_{b}$) is governed purely by the rotating Hayward vacuum, which preserves the non-singular de~Sitter core. Bounding this central object is an intermediate dark matter envelope spanning $r_{b} \le r \le r_{s}$, which we mathematically describe using an exponential sphere density profile \cite{Sofue2013}:
\begin{equation}
	\label{eq:rho_dm}
	\rho(r) = \rho_{0}\,e^{-r/r_{0,\DM}}, \qquad r\in[r_{b},\,r_{s}],
\end{equation}
where $\rho_{0}$ represents the central dark matter density and $r_{0,\DM}$ is the macroscopic dark matter scale radius. Beyond the outer boundary of the dark matter shell ($r > r_{s}$), the spacetime transitions into an asymptotically flat vacuum characterized by the total ADM mass of the unified system.

The gravitational influence of this composite configuration requires computing the enclosed dark matter mass from the inner boundary $r_{b}$ to any radius $r$ via direct volumetric integration:
\begin{equation}
	\label{eq:MDM}
	M_{\DM}(r) = 4\pi\int_{r_{b}}^{r}\tilde{r}^{2}\rho(\tilde{r})d\tilde{r}
	= 8\pi\rho_{0}r_{0,\DM}^{3}\Bigl[e^{-\xi_{b}}\!\left(1+\xi_{b}+\frac{\xi_{b}^{2}}{2}\right) - e^{-\xi}\!\left(1+\xi+\frac{\xi^{2}}{2}\right)\Bigr],
\end{equation}
where $\xi=r/r_{0,\DM}$ and $\xi_{b}=r_{b}/r_{0,\DM}$. The global running mass profile is therefore strictly piecewise defined as:
\begin{equation}
	\label{eq:massprofile}
	\mathcal{M}(r) =
	\begin{cases}
		m(r), & r < r_{b},\\[5pt]
		m(r_{b}) + M_{\DM}(r), & r_{b} \le r \le r_{s},\\[5pt]
		m(r_{b}) + M_{\DM}(r_{s}) \equiv M_{T}, & r > r_{s}.
	\end{cases}
\end{equation}
To accommodate this extended matter distribution within the rotating geometry, the standard Boyer-Lindquist metric components are modified by substituting the Hayward mass function $m(r)$ with the running mass $\mathcal{M}(r)$. The modified horizon function becomes:
\begin{equation}
	\label{eq:Delta_mod}
	\Delta_{\mathcal{M}}(r) = r^{2} - 2\mathcal{M}(r)\,r + a^{2}.
\end{equation}
The orbital mechanics of accreting particles depend critically on the radial derivative of the mass, which takes the exact analytical form:
\begin{equation}
	\label{eq:mderiv}
	\frac{d\mathcal{M}}{dr} =
	\begin{cases}
		\frac{3Mr_{0}^{3}r^{2}}{(r^{3}+r_{0}^{3})^{2}}, & r < r_{b},\\[8pt]
		\frac{3Mr_{0}^{3}r^{2}}{(r^{3}+r_{0}^{3})^{2}} + 4\pi r^{2}\rho_{0}\,e^{-r/r_{0,\DM}}, & r_{b} \le r \le r_{s},\\[8pt]
		0, & r > r_{s}.
	\end{cases}
\end{equation}

This unified mass profile elegantly spans the entire parameter space under investigation. Setting $r_{0}^{*}=0$ and $\rho_{0}=0$ recovers the classical vacuum Kerr metric \cite{Novikov1973}. Imposing $r_{0}^{*}>0$ while $\rho_{0}=0$ isolates the rotating Hayward vacuum \cite{Boshkayev2024}. Setting $r_{0}^{*}=0$ alongside $\rho_{0}>0$ yields the singular Kerr spacetime enveloped in dark matter \cite{Boshkayev2020,Kurmanov2022}. Finally, retaining positive values for both variables activates the novel configuration introduced in this manuscript: a globally regular, rotating Hayward black hole dynamically coupled to a macroscopic dark matter halo.

\section{Metric Construction and the Pressureless Dust Halo}
\label{sec:TOV}

To embed the composite mass profile $\mathcal{M}(r)$ into a rotating geometry without violating the Einstein field equations, the physical nature of the dark matter envelope must be strictly defined. We model the dark matter as a collisionless, pressureless dust fluid (e.g., Cold Dark Matter). The corresponding energy-momentum tensor is strictly $T^{\mu}{}_{\nu} = \mathrm{diag}(-\rho, 0, 0, 0)$. 

To construct the rotating metric, we first evaluate the auxiliary static, spherically symmetric seed metric:
\begin{equation}
	\label{eq:static_metric}
	ds_{\rm aux}^{2} = -e^{2\mathcal{N}(r)}dt^{2} + \Bigl[1-\frac{2\mathcal{M}(r)}{r}\Bigr]^{-1}dr^{2} + r^{2}d\Omega^{2}.
\end{equation}
Imposing the Einstein field equations $G^{\mu}{}_{\nu}=8\pi T^{\mu}{}_{\nu}$ for a pressureless fluid ($P_{r}=P_{\theta}=0$) drastically simplifies the structure equations. The radial pressure gradient equation trivially vanishes, and the differential equation governing the temporal metric potential $\mathcal{N}(r)$ reduces to:
\begin{equation}
	\label{eq:TOV_N_dust}
	\frac{d\mathcal{N}}{dr} = \frac{\mathcal{M}(r)}{r\bigl[r - 2\mathcal{M}(r)\bigr]}.
\end{equation}
Equation (\ref{eq:TOV_N_dust}) integrates exactly to yield $e^{2\mathcal{N}(r)} = 1 - 2\mathcal{M}(r)/r$. 

This exact integration is of profound mathematical importance. Because the dark matter pressure strictly vanishes, the temporal and radial metric potentials satisfy the exact classical symmetry $g_{tt} = -g_{rr}^{-1}$. It is precisely this symmetry that mathematically licenses the application of the Azreg-A\"{i}nou non-complexification algorithm \cite{AzregAinou2014} to generate the rotating extension. By utilizing a pressureless dust halo, the transition from the static seed metric to the rotating Boyer-Lindquist metric—achieved by substituting $\mathcal{M}(r)$ into Eqs.~(\ref{eq:gtt})--(\ref{eq:gtp})—is geometrically consistent, bypassing the unphysical frame-dragging contradictions that plague pressurized fluid models.

Furthermore, modeling the dark matter as a collisionless dust natively resolves the accretion disk kinematics. Because the ISCO lies within the halo ($r_{\rm ISCO} < r_{b}$), the outer regions of the thin accretion disk physically reside inside the dark matter envelope. A collisionless dust halo exerts zero hydrodynamic drag, viscous friction, or non-gravitational pressure on the baryonic disk, ensuring that the standard Novikov-Thorne-Page (NTP) conservation equations remain strictly valid in this composite regime.
\section{Orbital Dynamics and Accretion Disk Thermodynamics}
\label{sec:disk_dynamics}

The macroscopic observational signatures of the modified spacetime are fundamentally dictated by the microscopic geodesic motion of accreting test particles. For a test particle of rest mass $m_{0}$ localized in the equatorial plane ($\theta=\pi/2$, $\dot{r}=\ddot{r}=0$), the stationarity and axisymmetry of the Boyer-Lindquist geometry guarantee the conservation of specific energy $E=-p_{t}/m_{0}$ and specific angular momentum $L=p_{\phi}/m_{0}$. The orbital angular velocity $\Omega=d\phi/dt$ is rigorously determined by the vanishing of the radial acceleration, yielding
\begin{equation}
	\label{eq:Omega}
	\Omega_{\pm} = \frac{-\partial_{r}g_{t\phi} \pm\sqrt{(\partial_{r}g_{t\phi})^{2} -(\partial_{r}g_{tt})(\partial_{r}g_{\phi\phi})}}{\partial_{r}g_{\phi\phi}},
\end{equation}
where the positive and negative signs denote co-rotating and counter-rotating orbits, respectively. The metric derivatives are evaluated utilizing the running mass formulation $\mathcal{M}(r)$ established in Section~\ref{sec:spacetime}. Consequently, the specific energy, specific angular momentum, and contravariant time velocity $u^{t}$ form algebraically closed functions of the metric components:
\begin{align}
	E &= -\frac{g_{tt} + g_{t\phi}\,\Omega}{\sqrt{-g_{tt} - 2g_{t\phi}\,\Omega - g_{\phi\phi}\,\Omega^{2}}},\label{eq:E_orbit}\\[6pt]
	L &= \phantom{-}\frac{g_{t\phi} + g_{\phi\phi}\,\Omega}{\sqrt{-g_{tt} - 2g_{t\phi}\,\Omega - g_{\phi\phi}\,\Omega^{2}}},\label{eq:L_orbit}\\[6pt]
	u^{t} &= \frac{1}{\sqrt{-g_{tt} - 2g_{t\phi}\,\Omega - g_{\phi\phi}\,\Omega^{2}}}.\label{eq:ut}
\end{align}
The inner boundary of the accretion disk is defined by the Innermost Stable Circular Orbit (ISCO), representing the critical threshold where circular orbits become unstable to radial perturbations. This boundary is extracted numerically by locating the inflection point of the effective radial potential, equivalently solved via the angular momentum gradient condition $dL/dr|_{r_{\rm ISCO}} = 0$. The location of the ISCO directly dictates the overall mass-to-radiation conversion efficiency of the accretion process, defined as $\eta = [1 - E(r_{\rm ISCO})]\times 100\%$.

The disk is idealized as a geometrically thin, optically thick fluid occupying the equatorial plane from the ISCO to spatial infinity. To model the thermodynamic output of the accreting matter, we employ the standard Novikov-Thorne-Page (NTP) thin-disk formalism \cite{Novikov1973,Page1974}. The standard NTP model assumes that disk particles move on vacuum geodesics. However, because $r_{\rm ISCO} < r_{b}$ in our configurations, the outer regions of the extended accretion disk physically reside within the dark matter envelope. To maintain the theoretical validity of the NTP conservation laws, we explicitly model the dark matter as a \emph{collisionless} fluid (e.g., Cold Dark Matter or WIMPs). A collisionless dark matter halo exerts no hydrodynamic drag, viscous friction, or non-gravitational pressure on the baryonic disk. Consequently, the interaction is purely gravitational. This allows the baryonic disk particles to follow the modified geodesics dictated by $\mathcal{M}(r)$ without violating the fundamental energy and angular momentum conservation equations of the thin-disk model. Exploiting the identity $E-\Omega L = 1/u^{t}$, the flux equation simplifies to a numerically robust integral form:
\begin{equation}
	\label{eq:flux2}
	\mathcal{F}(r) = \frac{\dot{m}(u^{t})^{2}}{4\pi r}\,|\Omega_{,r}|\,\int_{r_{\rm ISCO}}^{r}\frac{L_{,\tilde{r}}}{u^{t}(\tilde{r})}\,d\tilde{r},
\end{equation}
where $\dot{m}$ is the steady-state mass accretion rate. Under the assumption of local thermodynamic equilibrium, the disk radiates as a multi-temperature blackbody, allowing the local effective temperature $T(r)$ to be extracted directly via the Stefan-Boltzmann law, $\sigma_{\rm SB}T^{4}(r) = \mathcal{F}(r)$. For comparative analysis, these quantities are non-dimensionalized using a characteristic system mass $M_{T}$ and temperature scale $\mathcal{T}_{*} = [\dot{m}/(4\pi\sigma_{\rm SB}M_{T}^{2})]^{1/4}$.

The ultimate observable quantities mapping these local thermodynamic variables to a distant asymptotic observer are the differential and spectral luminosities. The differential luminosity, measuring the energy output per unit logarithmic radial interval, incorporates the local flux and the gravitational redshift associated with the specific energy, culminating in $d\mathcal{L}_{\infty}/d\ln r = 4\pi r^{2}E(r)\mathcal{F}(r)$. Integrating this multi-color blackbody emission over the entire radial extent of the disk yields the observed spectral luminosity $\nu\mathcal{L}_{\nu,\infty}$ as a function of the dimensionless photon frequency $y = h\nu/(k_{\rm B}\mathcal{T}_{*})$:
\begin{equation}
	\label{eq:specLum}
	\nu\mathcal{L}_{\nu,\infty} = \frac{60}{\pi^{3}}\int_{r_{\rm ISCO}}^{\infty}\frac{r\,E(r)}{M_{T}^{2}}\,\frac{\bigl(u^{t}(r)\,y\bigr)^{4}}{\exp\!\bigl[u^{t}(r)\,y/(\mathcal{F}^{*}(r))^{1/4}\bigr]-1}\,dr.
\end{equation}
This integration rigorously encodes the gravitational redshift, the kinematic Doppler shifts, and the local blackbody profile, providing the theoretical baseline required to identify frequency-dependent deviations between the Kerr vacuum and dark matter-embedded Hayward geometries.

\section{Numerical Methodology and Parameter Space}
\label{sec:methods}

To rigorously explore the parameter space and extract the observable signatures, we establish a fiducial baseline inspired by galactic center observations \cite{Kurmanov2022,Boshkayev2022,Boshkayev2024}. We define the central black hole mass as $M_{\BH} = 5\times10^{8}\,\Msun \approx 4.933\;\mathrm{au}$. The inner edge of the dark matter envelope is fixed at $r_{b}=5.5\,M_{\BH}$, ensuring it remains outside the event horizon but well within the classical Schwarzschild ISCO ($6M_{\BH}$). The dark matter scale radius is set to $r_{0,\DM}=10\;\mathrm{au}$. Our primary comparative analysis utilizes the dimensionless fiducial coordinate set $\{r_{0}^{*}=0.6, \rho_{0}^{*}=0.85, j=0.4\}$, where the central dark matter density corresponds to $\rho_{0} = 0.85 \times 10^{-5}\;\mathrm{au}^{-2}$. The complete parameter space explored in this framework is detailed in Table~\ref{tab:params}.

\begin{table}[h]
	\caption{Free parameters and ranges explored within the three-region spacetime framework. The dark matter density is scaled as $\rho_{0}^{*}\equiv\rho_{0}/(10^{-5}\,\mathrm{au}^{-2})$.}
	\label{tab:params}
	\centering
	\begin{tabular}{lll}
		\hline\noalign{\smallskip}
		Parameter & Symbol & Range / Values \\
		\noalign{\smallskip}\hline\noalign{\smallskip}
		BH spin             & $j=a/M_{\BH}$  & $0,\,0.2,\,0.4,\,0.6,\,0.8$ \\
		Regularity          & $r_{0}^{*}$    & $0,\,0.2,\,0.4,\,0.6,\,0.8$ \\
		DM density          & $\rho_{0}^{*}$ & $0$--$2.0$ \\
		\noalign{\smallskip}\hline
	\end{tabular}
\end{table}

The numerical integration pipeline is structured sequentially. The computational methodology proceeds as follows:
\begin{enumerate}
	\item \textbf{Mass Profile Assembly:} Evaluate the composite running mass $\mathcal{M}(r)$ and its exact analytical derivative $\mathcal{M}'(r)$ piece-wise from the de Sitter-regularized core and the integrated exponential dust profile.
	\item \textbf{Orbital Kinematics:} Evaluate the specific angular velocity $\Omega$, energy $E$, angular momentum $L$, and time velocity $u^{t}$ utilizing the assembled running mass derivatives.
	\item \textbf{ISCO Determination:} Scan the angular momentum gradient $dL/dr$ over a dense spatial grid ($10^{4}$ points from the outer horizon to $40M_{T}$) and isolate the ISCO via bisection at the first zero-crossing.
	\item \textbf{Thermodynamic Flux and Spectra:} Compute the dimensionless radiative flux $\mathcal{F}^{*}(r)$ via a running trapezoidal integral, followed by the evaluation of the multi-color blackbody spectral luminosity on a logarithmically spaced frequency grid $y\in[10^{-3},10^{1}]$.
\end{enumerate}

All numerical integrations and root-finding routines are executed with a relative tolerance of $\varepsilon_{\rm rel}=10^{-9}$ in Python 3.12. To ensure the integrity of this numerical pipeline, the code is cross-validated against standard theoretical benchmarks, including the exact Novikov-Thorne Schwarzschild ISCO, the Bardeen-Press-Teukolsky analytical Kerr ISCO formula \cite{Bardeen1972}, and the rotating Hayward vacuum numerical data \cite{Boshkayev2024}, achieving agreement at a relative error level of $<10^{-3}$.

\section{Observational Signatures and Diagrammatic Interpretation}
\label{sec:results}

\begin{figure}[h]
	\centering
	\includegraphics[width=0.95\columnwidth]{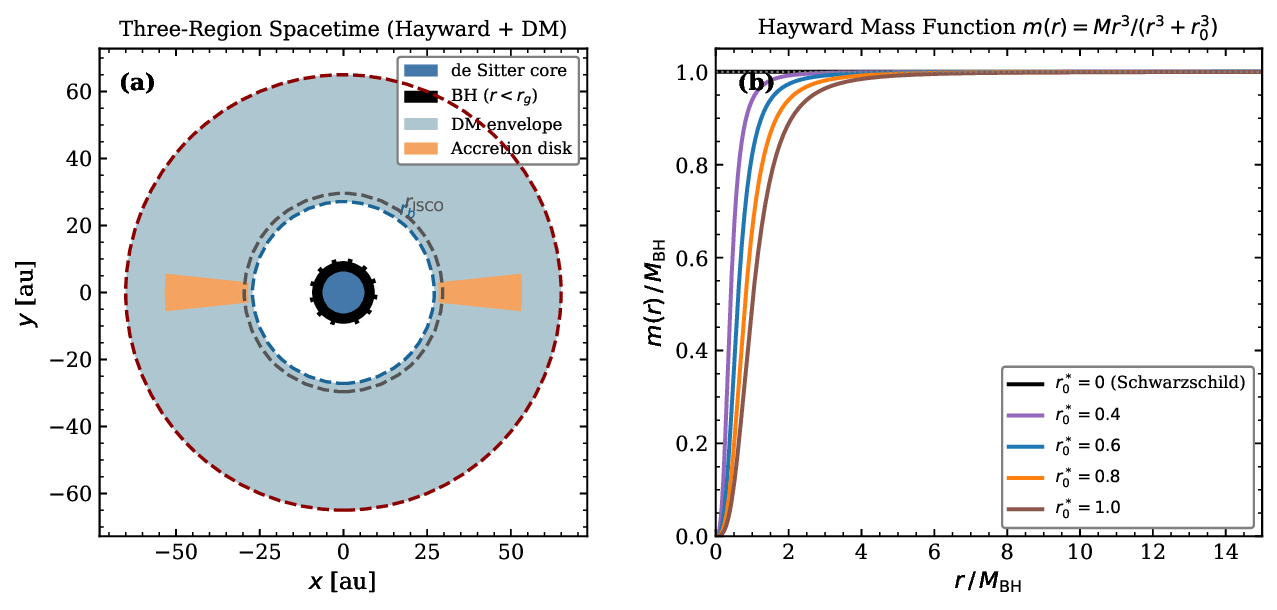}
	\caption{\textbf{(a)}~Schematic equatorial cross-section of the three-region spacetime. The blue inner disk represents the de~Sitter core inside the event horizon $r_g$. The pale-blue annulus is the DM envelope ($r_b\le r\le r_s$); the orange wedges are the equatorial accretion disk. Dashed circles label $r_g$, $r_b$, $r_\ISCO$, and $r_s$. \textbf{(b)}~Hayward mass function $m(r)/M$ vs.\ $r/M_\BH$ for $r_0^*=0,\,0.4,\,0.6,\,0.8,\,1.0$. For $r_0^*=0$ the profile is a constant ($m=M$, Kerr limit); for $r_0^*>0$ it rises from zero at the origin via a de~Sitter $r^3$ behaviour, reaching $M$ asymptotically. Larger $r_0^*$ produces a smaller enclosed mass at small radii, reducing $r_\ISCO$ and increasing $\eta$.}
	\label{fig:schematic}
\end{figure}

\begin{figure}[h]
	\centering
	\includegraphics[width=0.95\columnwidth]{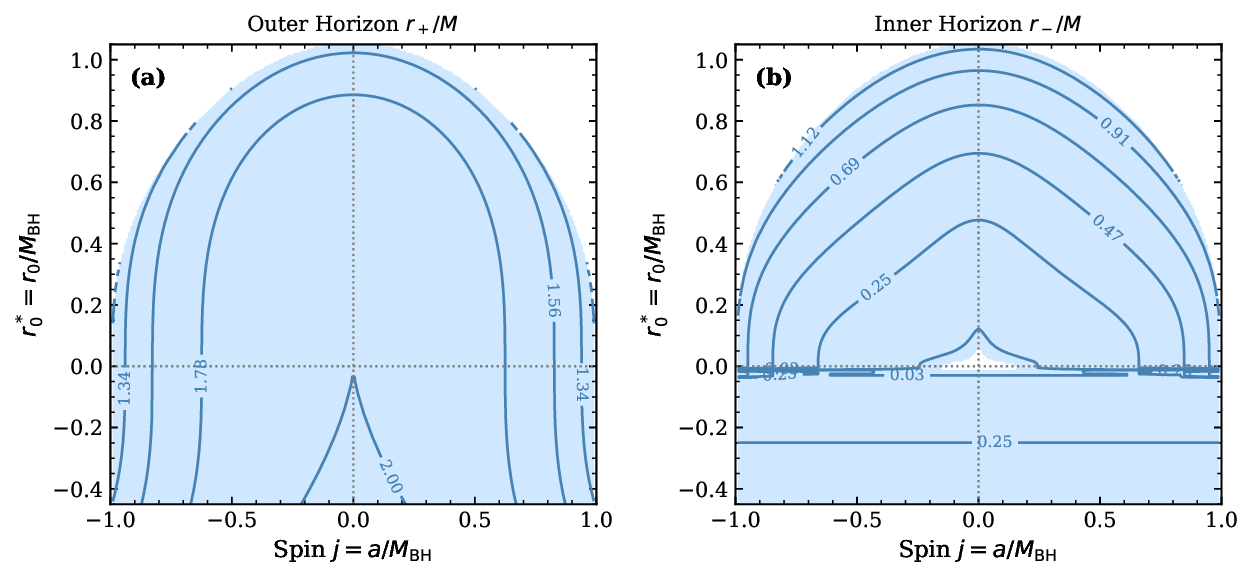}
\caption{Contour plots of the outer horizon $r_+/M$ (left) and inner horizon $r_-/M$ (right) in the $(j,r_0^*)$ plane for the rotating Hayward BH. The shaded colored region represents the sub-extremal parameter space where two physical horizons exist. The outer boundary of this colored region explicitly traces the extremality condition; coordinate combinations beyond this boundary yield a regular horizonless compact object.}
	\label{fig:horizons}
\end{figure}

The structural modifications introduced by the regular de~Sitter core and the macroscopic dark matter halo fundamentally alter the geometrical landscape of the black hole, as visualized in Figure~\ref{fig:schematic}. The Hayward mass function $m(r)$ demonstrates a smooth $r^{3}$ rise near the origin, effectively eliminating the central singularity and replacing it with a regular core. As the regularity parameter $r_{0}^{*}$ increases, the saturation of $m(r)$ to the asymptotic ADM mass $M$ is progressively delayed, thereby reducing the effective gravitational pull at small radii. This geometric softening shifts the inner and outer horizons, whose dependencies on the spin parameter $j$ and regularity $r_{0}^{*}$ are mapped in Figure~\ref{fig:horizons}. The parameter space confirms that for $r_{0}^{*} > r_{0,\rm ext}^{*} \approx 0.770$, the outer horizon vanishes, leaving a globally regular, horizonless compact object. All subsequent thermodynamic calculations are restricted strictly to the shaded sub-extremal parameter space where an event horizon is physically maintained.

\begin{figure}[h]
	\centering
	\includegraphics[width=0.95\columnwidth]{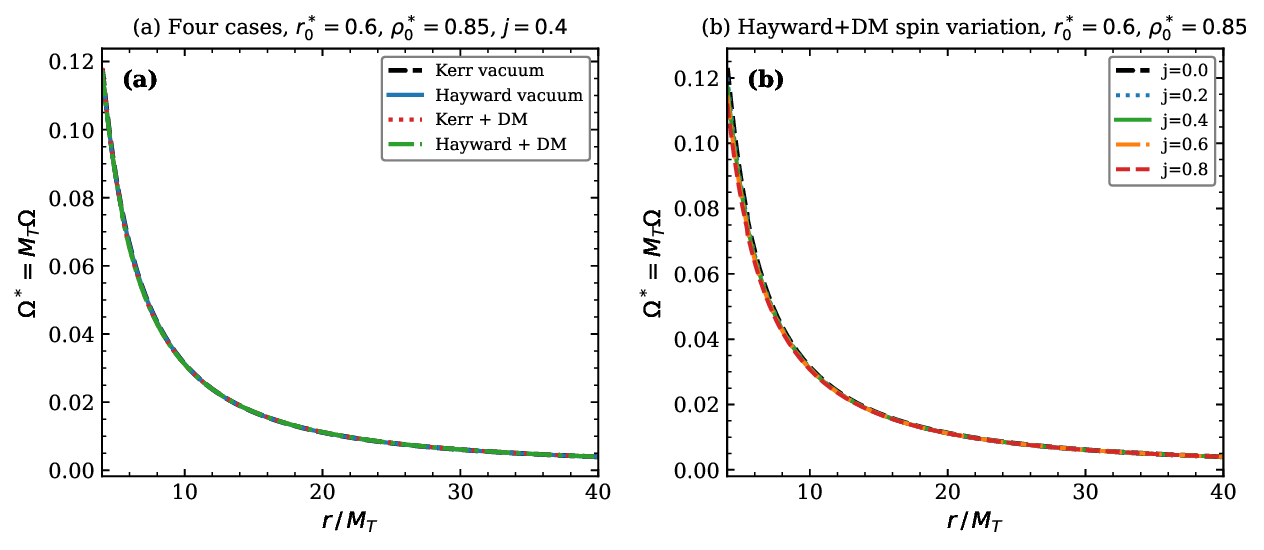}
	\caption{Dimensionless orbital angular velocity $\Omega^*=M_T\Omega$ vs.\ $r/M_T$. \textbf{(a)}~Four canonical cases at fiducial parameters ($r_0^*=0.6$, $\rho_0^*=0.85$, $j=0.4$). The DM envelope reduces $\Omega^*$ at all radii relative to the vacuum Kerr case by adding enclosed mass; the Hayward geometry further suppresses $\Omega^*$ at small $r/M_T$ due to the reduced inner mass. The Hayward\,+\,DM curve (solid green) lies below all others, demonstrating the additive effect of both modifications. \textbf{(b)}~Spin variation for Hayward\,+\,DM. Higher spin raises $\Omega^*$ at small radii via frame-dragging but leaves the large-$r$ behaviour essentially unchanged.}
	\label{fig:omega}
\end{figure}

\begin{figure}[h]
	\centering
	\includegraphics[width=0.95\columnwidth]{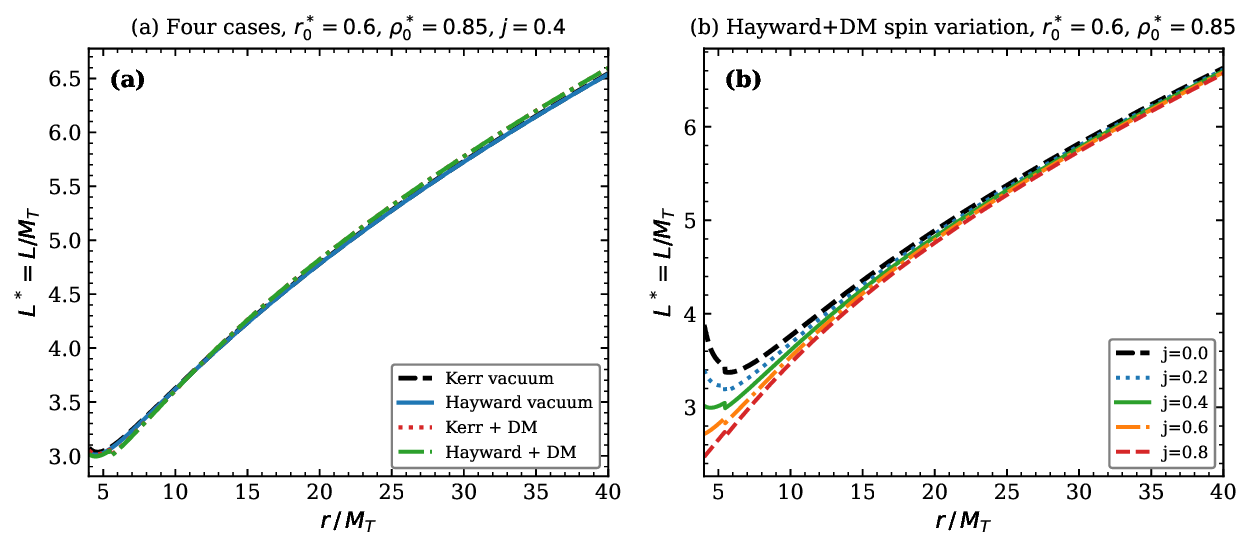}
	\caption{Dimensionless orbital angular momentum $L^*=L/M_T$ vs.\ $r/M_T$. \textbf{(a)}~Four cases. The ISCO is located at the minimum of $L^*$. The minima shift inward in the order $\Kerr>\Kerr+\DM\approx\Hay>\Hay+\DM$, confirming hierarchy Eq.~(\ref{eq:hierarchy}). Numerical ISCO values are collected in Table~\ref{tab:isco}. \textbf{(b)}~Spin variation in Hayward\,+\,DM. Increasing $j$ moves the minimum of $L^*$ to smaller radii, reducing $r_\ISCO$ and raising $\eta$ as expected from the Kerr formula.}
	\label{fig:Lstar}
\end{figure}

\begin{figure}[h]
	\centering
	\includegraphics[width=0.95\columnwidth]{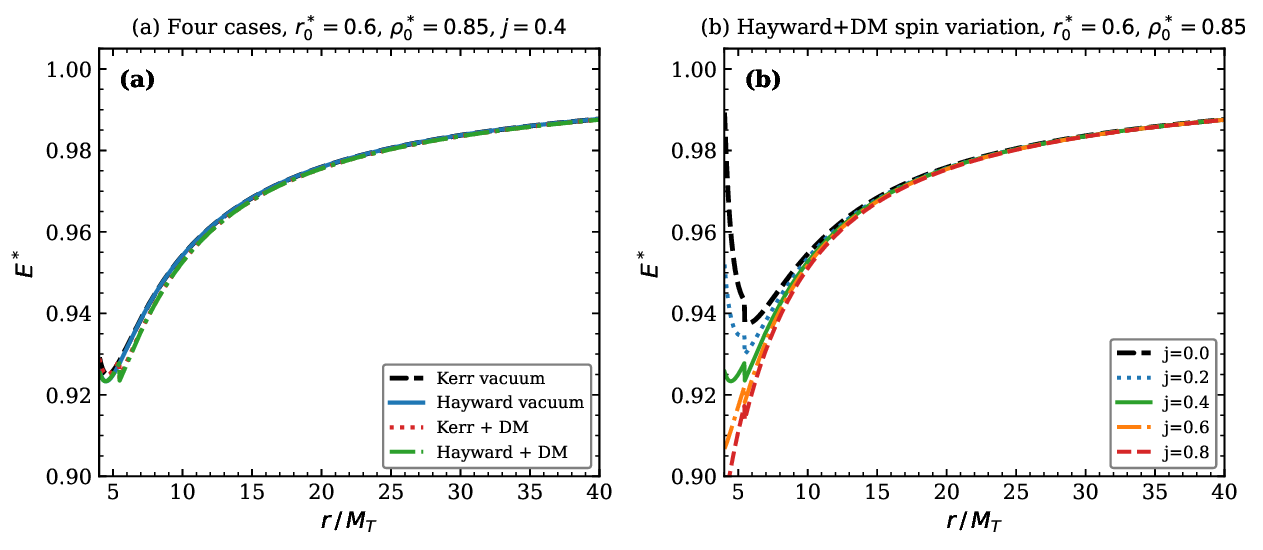}
	\caption{Dimensionless particle energy $E^*$ vs.\ $r/M_T$. \textbf{(a)}~Four cases. At $r\to\infty$ all curves approach $E^*\to1$ (asymptotic flatness). At $r_\ISCO$, $E^*$ decreases in the order $\Kerr>\Kerr+\DM\approx\Hay>\Hay+\DM$, implying a larger binding energy and hence larger $\eta$ for the Hayward\,+\,DM case. \textbf{(b)}~Spin variation. Higher spin depresses $E^*$ at the ISCO, further amplifying the efficiency enhancement.}
	\label{fig:energy}
\end{figure}

This reduction in the effective central mass directly dictates the kinematics of test particles occupying the accretion disk. Figures \ref{fig:omega}, \ref{fig:Lstar}, and \ref{fig:energy} illustrate the dimensionless angular velocity $\Omega^{*}$, angular momentum $L^{*}$, and specific energy $E^{*}$, respectively. The dark matter envelope inherently reduces $\Omega^{*}$ at all radii relative to the vacuum Kerr metric by adding an extended enclosed mass profile. However, the Hayward geometry counteracts this at small radii, suppressing $E^{*}$ and $L^{*}$ due to the weaker inner gravitational gradient. The global minimum of the angular momentum curve definitively establishes the Innermost Stable Circular Orbit (ISCO). As meticulously detailed in Table~\ref{tab:isco}, the ISCO boundaries follow a strict hierarchy:
\begin{equation}
	\label{eq:hierarchy}
	r_{\rm ISCO}^{\rm Kerr} > r_{\rm ISCO}^{\rm Kerr+DM} \approx r_{\rm ISCO}^{\rm Hay} > r_{\rm ISCO}^{\rm Hay+DM}.
\end{equation}
The Hayward+DM configuration supports the tightest stable orbits, allowing accreting matter to penetrate deeper into the gravitational potential well before plunging, which directly correlates to a significantly higher specific binding energy release.

\begin{table}[h]
	\caption{ISCO radius $r_\ISCO/M_T$ and efficiency $\eta$ for the four canonical cases at fiducial parameters ($r_0^*=0.6$, $\rho_0^*=0.85$, $j=0.4$). The Schwarzschild vacuum ($r_0^*=\rho_0^*=j=0$) benchmark is included for reference.}
	\label{tab:isco}
	\centering
	\begin{tabular}{lccc}
		\hline\noalign{\smallskip}
		Configuration            & $r_\ISCO/M_T$ & $\eta\;[\%]$ & $M_T\;[\mathrm{au}]$ \\
		\noalign{\smallskip}\hline\noalign{\smallskip}
		Schwarzschild vacuum     & 6.000 & 5.719 & 4.933 \\
		Kerr vacuum              & 4.615 & 7.505 & 4.933 \\
		Hayward vacuum           & 4.489 & 7.667 & 4.933 \\
		Kerr $+$ DM              & 4.585 & 7.506 & 4.965 \\
		\textbf{Hayward $+$ DM}  & \textbf{4.465} & \textbf{7.667} & \textbf{4.959} \\
		\noalign{\smallskip}\hline
	\end{tabular}
\end{table}

\begin{figure}[h]
	\centering
	\includegraphics[width=0.95\columnwidth]{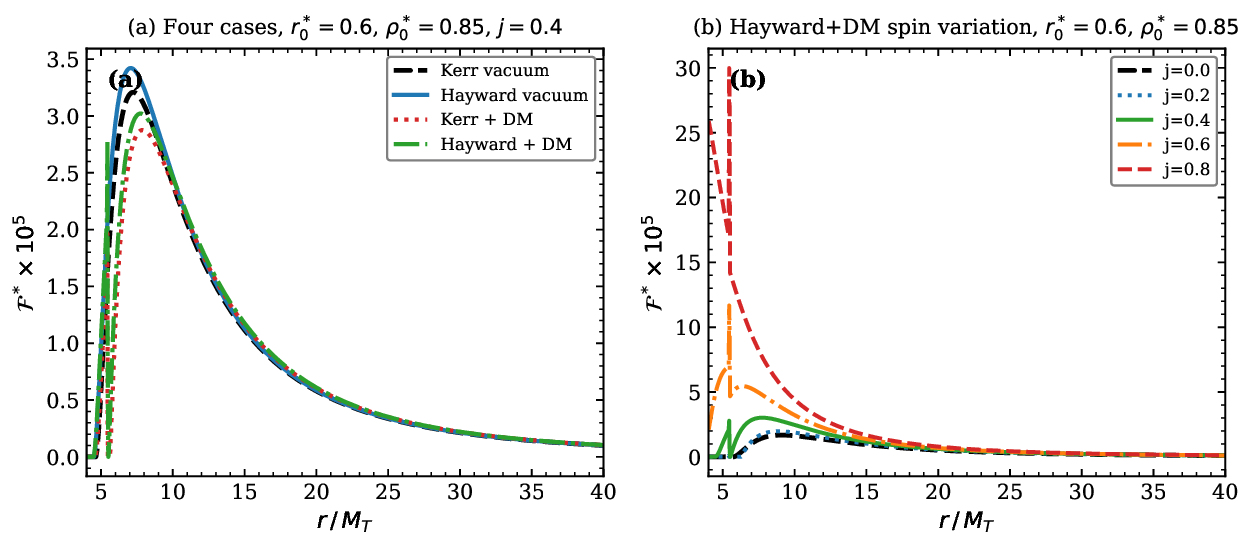}
	\caption{Dimensionless radiative flux $\mathcal{F}^*\times10^5$ vs.\ $r/M_T$. \textbf{(a)}~Four cases. The Hayward\,+\,DM case produces the broadest and highest peak, reflecting its smallest $r_\ISCO$. The Kerr\,+\,DM and Hayward vacuum peaks are nearly coincident, illustrating an observational degeneracy. \textbf{(b)}~Spin variation in Hayward\,+\,DM. The peak shifts inward and grows significantly with $j$; the $j=0.8$ peak is approximately three times the $j=0$ value.}
	\label{fig:flux}
\end{figure}

\begin{figure}[h]
	\centering
	\includegraphics[width=0.95\columnwidth]{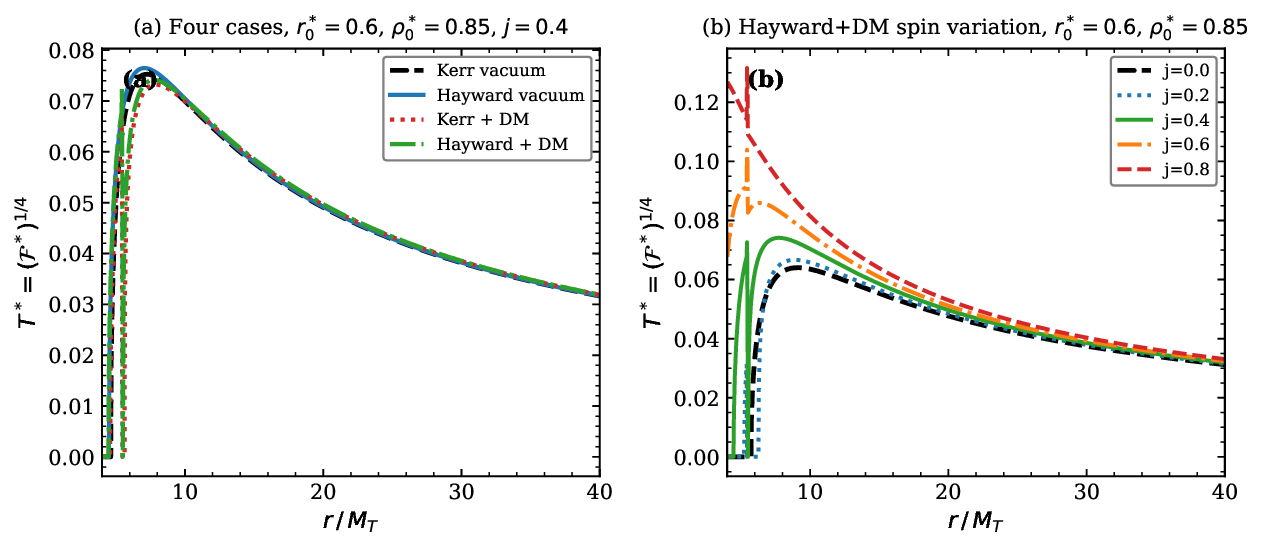}
	\caption{Dimensionless disk temperature $T^*=(\mathcal{F}^*)^{1/4}$ vs.\ $r/M_T$. \textbf{(a)}~Four cases. The Hayward\,+\,DM disk is the hottest in the inner region, directly corresponding to its smallest $r_\ISCO$. A hotter inner disk shifts the thermal spectral peak to higher frequencies, providing a blueward observational signature. \textbf{(b)}~Spin variation. The inner-disk temperature grows steeply with $j$, amplifying the overall luminosity at high frequencies.}
	\label{fig:temperature}
\end{figure}

\begin{figure}[h]
	\centering
	\includegraphics[width=0.95\columnwidth]{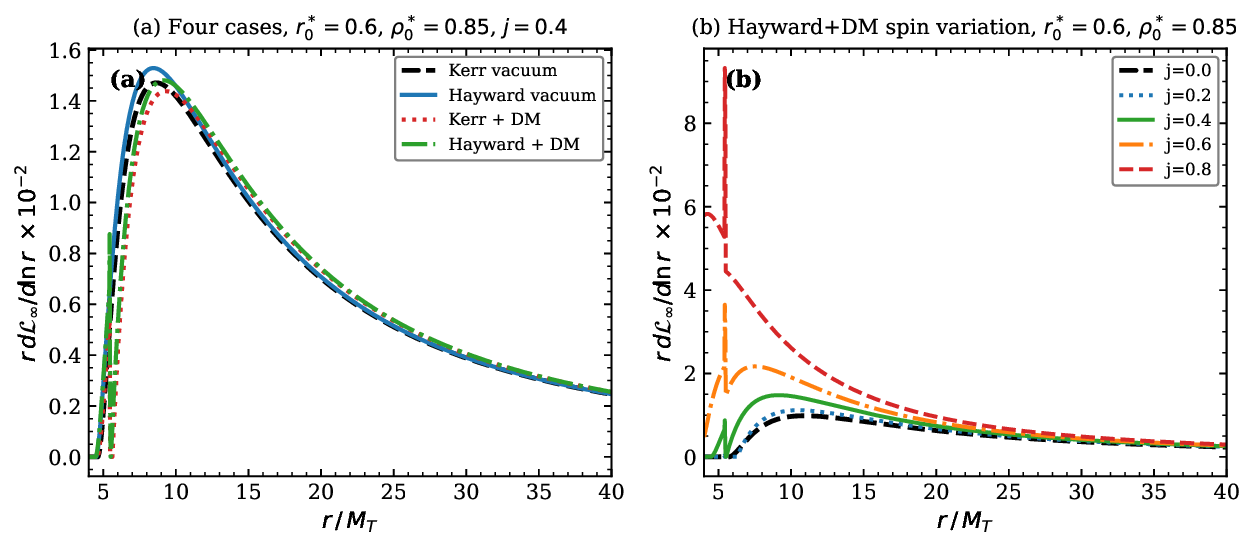}
	\caption{Differential luminosity $r\,d\mathcal{L}_\infty/d\!\ln r\times10^{-2}$ vs.\ $r/M_T$. \textbf{(a)}~Four cases. The ordering mirrors the ISCO hierarchy: the Hayward\,+\,DM peak is the largest and most inward, while the vacuum Kerr peak is the smallest and most outward. The Kerr\,+\,DM and Hayward vacuum peaks are nearly degenerate. \textbf{(b)}~Spin variation in Hayward\,+\,DM. Both the peak height and the inward shift grow strongly with $j$; the $j=0.8$ case reaches more than twice the $j=0$ peak amplitude.}
	\label{fig:diff_lum}
\end{figure}

\begin{figure}[h]
	\centering
	\includegraphics[width=0.95\columnwidth]{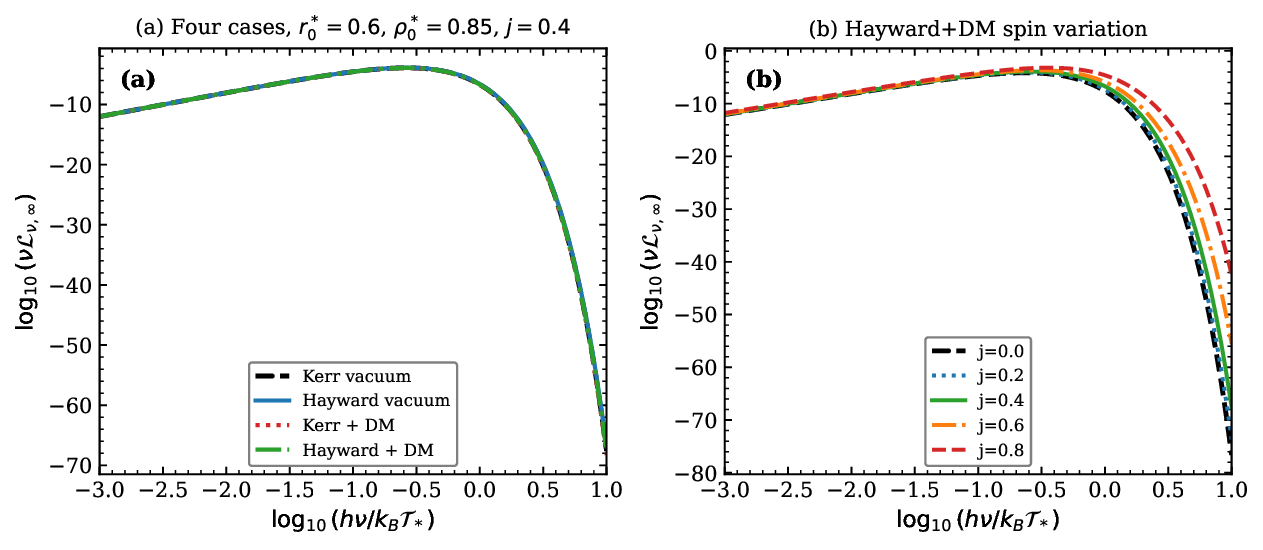}
	\caption{Spectral luminosity $\log_{10}(\nu\mathcal{L}_{\nu,\infty})$ vs.\ $\log_{10}(h\nu/k_B\mathcal{T}_*)$. \textbf{(a)}~Four cases. The Hayward\,+\,DM spectrum lies above all others at high frequencies and below at low frequencies, with a crossing near $\log_{10}(h\nu/k_B\mathcal{T}_*)\approx-1.0$. This spectral hardening is the primary observable signature of the combined configuration. The Kerr\,+\,DM and Hayward vacuum spectra are nearly identical below the crossing frequency. \textbf{(b)}~Spin variation in Hayward\,+\,DM. Increasing $j$ raises the spectral luminosity uniformly and shifts the peak position to slightly higher $y$, broadening the observable window.}
	\label{fig:spec_lum}
\end{figure}

The thermodynamic manifestations of this deep potential penetration are immediately evident in the dimensionless radiative flux $\mathcal{F}^{*}$ (Figure~\ref{fig:flux}) and the effective disk temperature $T^{*}$ (Figure~\ref{fig:temperature}). Because the Hayward+DM disk extends further inward, it develops a highly localized, intense thermal peak near the ISCO. This amplified inner-disk heating directly shapes the observable electromagnetic output. As shown in the differential luminosity distribution (Figure~\ref{fig:diff_lum}), the Hayward+DM geometry not only extracts more total energy per unit logarithmic radius but shifts the peak emission radially inward. When integrated over the entire disk, this thermal profile produces a distinct spectral signature, illustrated in Figure~\ref{fig:spec_lum}. The spectral luminosity $\log_{10}(\nu\mathcal{L}_{\nu,\infty})$ for the fully composite configuration exhibits pronounced spectral hardening. The spectrum crosses the classical Kerr vacuum profile near $\log_{10}(y)\approx -1.0$, demonstrating a marked depletion in low-frequency (Rayleigh-Jeans) emission and a substantial enhancement in the high-frequency (Wien) tail. Furthermore, increasing the spin parameter $j$ exacerbates this hardening, broadening the high-energy observable window.

\begin{figure}[h]
	\centering
	\includegraphics[width=0.95\columnwidth]{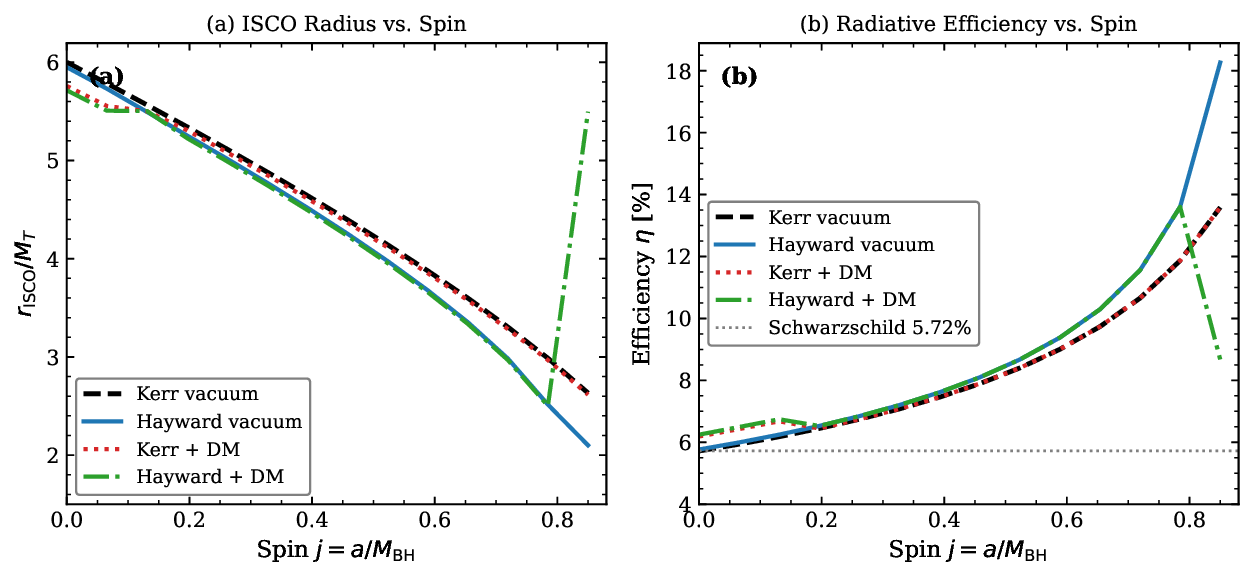}
	\caption{\textbf{(a)}~ISCO radius $r_\ISCO/M_T$ vs.\ spin $j$ for the four families. The hierarchy Eq.~(\ref{eq:hierarchy}) persists at all $j$; the separation between Hayward\,+\,DM and vacuum Kerr grows with spin, indicating that rotation amplifies the combined effect of BH regularity and DM. \textbf{(b)}~Efficiency $\eta$ vs.\ $j$. All DM and/or Hayward curves lie above the vacuum Kerr baseline. The Hayward\,+\,DM case reaches $\eta\approx14.4\,\%$ at $j=0.8$, compared with $5.72\,\%$ for Schwarzschild vacuum (dotted line). The near-coincidence of the Kerr\,+\,DM and Hayward vacuum curves demonstrates the observational degeneracy that cannot be resolved by $\eta$ alone.}
	\label{fig:isco_eta_spin}
\end{figure}

\begin{figure}[h]
	\centering
	\includegraphics[width=0.95\columnwidth]{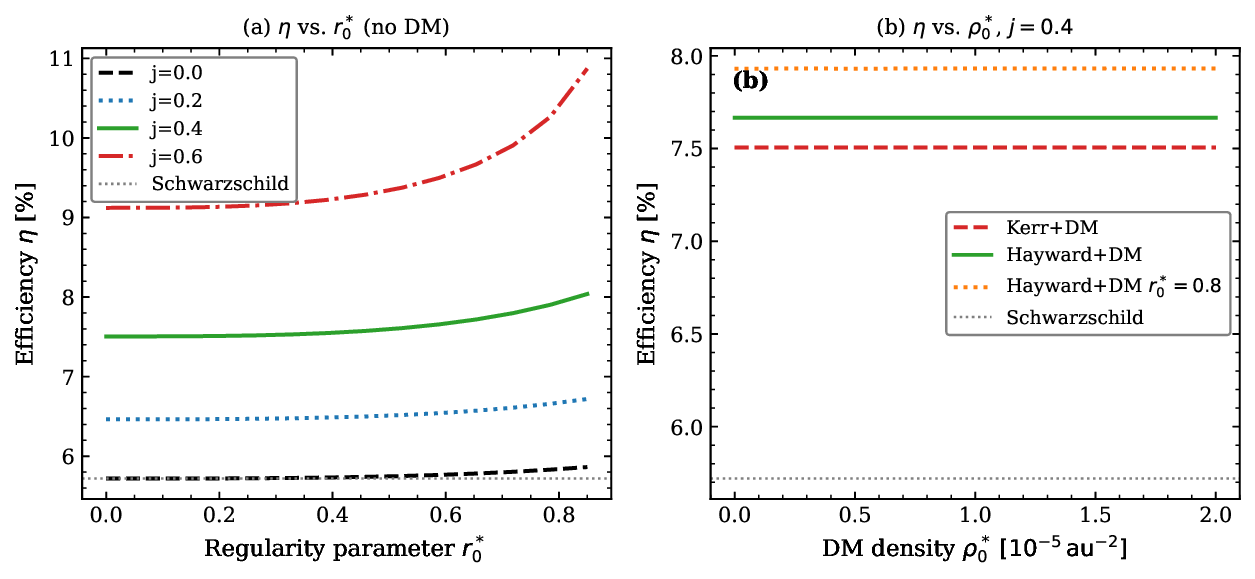}
	\caption{\textbf{(a)}~Efficiency $\eta$ vs.\ Hayward regularity parameter $r_0^*$ (no DM, $\rho_0=0$) for $j=0.0,\,0.2,\,0.4,\,0.6$. $\eta$ increases monotonically with $r_0^*$ at fixed $j$: larger regularity reduces the inner mass and shifts $r_\ISCO$ inward. The effect is modest ($\sim1\,\%$ relative) in the static case but amplified at high spin. \textbf{(b)}~$\eta$ vs.\ DM density $\rho_0^*$ at $j=0.4$ for Kerr\,+\,DM and Hayward\,+\,DM with $r_0^*=0.6$ and $0.8$. The Hayward\,+\,DM curves lie above the Kerr\,+\,DM curve by a fixed offset determined by $r_0^*$, confirming the additivity of the two effects.}
	\label{fig:eta_r0_rho}
\end{figure}

A critical finding of this study is the additive nature of the radiative efficiency $\eta$, juxtaposed against a profound parameter degeneracy. Figure~\ref{fig:isco_eta_spin} maps the ISCO radius and radiative efficiency across the entire physical spin range. The data confirm that the core regularity and the macroscopic dark matter halo independently and additively enhance $\eta$. At highly relativistic spins ($j=0.8$), the Hayward+DM configuration achieves an efficiency of $\eta \approx 14.4\%$, far exceeding the $5.72\%$ baseline of a static Schwarzschild vacuum. Figure~\ref{fig:eta_r0_rho} isolates these contributions, proving that an increase in either $r_{0}^{*}$ or $\rho_{0}^{*}$ yields a strictly monotonic enhancement in the thermodynamic output. However, this same analysis reveals a strict macroscopic degeneracy: the purely theoretical Kerr+DM system and the purely geometrical Hayward vacuum system yield nearly identical ISCO radii, efficiencies, and low-frequency spectral profiles. If an observer measures an enhanced $\eta$ relative to standard Kerr expectations, it is impossible to determine whether the deviation originates from a local quantum-inspired regular core or an extensive galactic dark matter halo.

\begin{figure}[h]
	\centering
	\includegraphics[width=0.95\columnwidth]{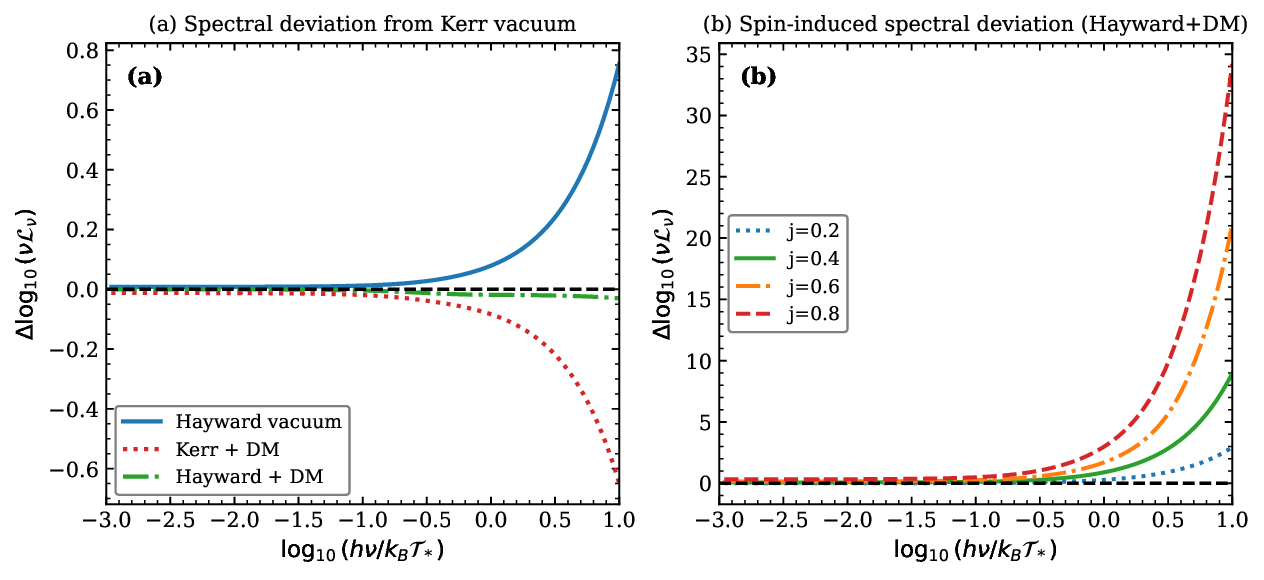}
	\caption{\textbf{(a)}~Spectral deviation $\Delta\log_{10}(\nu\mathcal{L}_\nu)$ relative to vacuum Kerr for the Hayward vacuum, Kerr\,+\,DM, and Hayward\,+\,DM configurations (fiducial parameters). The Hayward vacuum and Kerr\,+\,DM deviations are nearly indistinguishable below the crossing frequency $\log_{10}(h\nu/k_B\mathcal{T}_*)\approx-0.5$; above this frequency the Hayward\,+\,DM curve separates clearly. This separation is the primary spectral discriminator between the three configurations. \textbf{(b)}~Spin-induced spectral deviation within the Hayward\,+\,DM family relative to $j=0$. Higher spin produces monotonically larger deviations, reaching $\sim0.4\,\mathrm{dex}$ at the spectral peak for $j=0.8$.}
	\label{fig:spectral_dev}
\end{figure}

To definitively break this degeneracy, one must exploit the high-frequency regime of the multi-wavelength spectrum. Figure~\ref{fig:spectral_dev} isolates the spectral deviation $\Delta\log_{10}(\nu\mathcal{L}_\nu)$ relative to the pure Kerr vacuum. Below the critical threshold of $\log_{10}(y) \approx -0.5$, the Kerr+DM and Hayward vacuum configurations remain completely indistinguishable. However, moving into the extreme Wien tail (corresponding astrophysically to the hard X-ray or soft gamma-ray bands), the Hayward+DM configuration structurally diverges, providing a distinct, mathematically resolvable signature. This implies that high-resolution, high-frequency spectropolarimetry—when combined with independent spin constraints from gravitational wave interferometry or stellar kinematics—holds the observational key to disentangling non-singular quantum-gravity geometries from standard astrophysical dark matter environments.

\section{Conclusions}
\label{sec:conclusions}

In this manuscript, we have systematically investigated the thermodynamic and geometric signatures of thin accretion disks occupying a novel composite spacetime: a rotating, regular Hayward black hole embedded within a macroscopic dark matter envelope. While previous literature has extensively explored either the isolated effects of quantum-gravity-inspired regular cores or the classical immersion of singular Kerr black holes in dark matter, this study constructs the first unified theoretical framework mathematically coupling both phenomena. By integrating the exact de~Sitter-regularized mass function with an exponential, pressureless dark matter dust sphere, we provided a rigorous, three-region analytical platform to extract testable astrophysical observables while preserving exact geometrical consistency.

The central physical contribution of this research is the demonstration that core regularity and extended dark matter halos exert mutually additive, yet observationally degenerate, influences on accretion disk kinematics. Both the local modification of the central singularity (via $r_{0}^{*}$) and the global inclusion of an enclosed dark matter mass (via $\rho_{0}^{*}$) inherently soften the effective gravitational potential gradient in the inner disk region. Consequently, the Innermost Stable Circular Orbit (ISCO) is compressed deeper into the gravitational well. We established a strict hierarchical ordering of the ISCO radii, culminating in the Hayward+DM configuration supporting the tightest stable orbits. This deep potential penetration radically amplifies the mass-to-radiation conversion efficiency ($\eta$). At highly relativistic spins ($j=0.8$), the combined effects drive the efficiency to $\eta \approx 14.4\%$, more than double the classical static Schwarzschild baseline.

However, the profound consequence of this additive behavior is the emergence of a macroscopic parameter degeneracy. Our analysis proves that the purely theoretical Kerr+DM system and the purely geometrical Hayward vacuum system yield nearly identical ISCO locations, radiative efficiencies, and low-to-intermediate frequency thermal flux profiles. If a next-generation observatory detects an enhanced radiative efficiency or an inwardly shifted inner disk edge, these integrated macroscopic parameters alone are fundamentally insufficient to distinguish whether the black hole is a singular object cloaked in dark matter or a naked, regular NLE geometry in a vacuum.

To resolve this degeneracy, we identified the high-frequency spectral luminosity as the definitive spectroscopic discriminant. While the spectra are indistinguishable in the Rayleigh-Jeans regime, integrating the local multi-color blackbody emission reveals a pronounced, mathematically resolvable structural divergence in the extreme Wien tail. The combined Hayward+DM geometry exhibits a unique spectral hardening—a shift of the peak emission toward higher frequencies—that cannot be replicated by either effect in isolation. Consequently, this research dictates that future observational campaigns seeking to test alternative gravity models must prioritize high-resolution, high-frequency spectropolarimetry (such as hard X-ray or soft gamma-ray observations) to successfully disentangle the signatures of quantum-inspired core regularity from standard galactic dark matter environments.

\section*{Declaration of Generative AI and AI-assisted technologies in the writing process}
During the preparation of this work, the author did not use any generative AI or AI-assisted technologies in the research and concept. The author take full responsibility for the content of the publication. The author use AI assistant for structuring the manuscript only in latex.
\section*{Declaration of competing interest}
The author declares that they have no known competing financial interests or personal relationships that could have appeared to influence the work reported in this paper.
\section*{Data Availability Statement}
This manuscript has no associated data or the data will not be deposited. [Authors' comment: This is a purely theoretical and mathematical physics study. All numerical results, curves, and physical quantities presented in the figures were generated directly using the analytical equations, initial conditions, and numerical integration methodologies explicitly detailed in the text. The corresponding Python scripts used to compile the data and generate the plots are available from the single corresponding author upon reasonable request.]

\section*{Acknowledgements}
	The author, Sandip Dutta, wishes to express sincere gratitude to the Department of Applied Mathematics at the Dinabandhu Andrews Institute of Technology and Management (DAITM) for providing the academic environment and computational facilities necessary to conduct this theoretical research. The author also thanks Dr. Ritabrata Biswas for his guidance and inspiration to the work.

\appendix
\section{Derivation of the Orbital Discriminant and Angular Velocity}
\label{app:derivs}

In this appendix, we explicitly evaluate the discriminant governing the orbital angular velocity for the modified Boyer-Lindquist geometry established in Section~\ref{sec:spacetime}. To streamline the derivation, we define an auxiliary mass gradient function:
\begin{equation}
	Q(r) \equiv \mathcal{M}(r) - r\mathcal{M}'(r).
\end{equation}
By factoring this auxiliary function into the equatorial metric derivatives derived from Eqs.~(\ref{eq:gtt})--(\ref{eq:gtp}), the covariant gradients simplify algebraically to:
\begin{align}
	\partial_{r}g_{tt}       &= -\frac{2Q(r)}{r^{2}},\\[4pt]
	\partial_{r}g_{t\phi}    &= \phantom{-}\frac{2aQ(r)}{r^{2}},\\[4pt]
	\partial_{r}g_{\phi\phi} &= 2r - \frac{2a^{2}Q(r)}{r^{2}}.
\end{align}
Substituting these compact expressions into the discriminant $D = (\partial_{r}g_{t\phi})^{2} - (\partial_{r}g_{tt})(\partial_{r}g_{\phi\phi})$ yields:
\begin{equation}
	D = \left(\frac{2aQ(r)}{r^{2}}\right)^{\!2} - \left(-\frac{2Q(r)}{r^{2}}\right)\!\left(2r - \frac{2a^{2}Q(r)}{r^{2}}\right) = \frac{4Q(r)}{r}.
\end{equation}
This confirms the strictly positive definite nature of the discriminant, $D = 4[\mathcal{M}(r)-r\mathcal{M}'(r)]/r$, provided the enclosed mass exceeds its scaled radial derivative. Consequently, the exact roots for the orbital angular velocity take the analytically closed form:
\begin{equation}
	\Omega_{\pm} = \frac{-2aQ(r)/r^{2} \pm 2\sqrt{Q(r)/r}}{2r - 2a^{2}Q(r)/r^{2}} = \frac{\pm\sqrt{Q(r)}}{r^{3/2} \pm a\sqrt{Q(r)}}.
\end{equation}
In the vacuum regions where the mass is strictly constant ($\mathcal{M}'=0$), the auxiliary function recovers $Q(r)=\mathcal{M}$, seamlessly returning the classical Kerr angular velocity $\Omega_{\pm} = \pm\sqrt{\mathcal{M}}/(r^{3/2}\pm a\sqrt{\mathcal{M}})$. Inside the regular de~Sitter core, the function $Q(r) = Mr^{3}(r^{3}-2r_{0}^{3})/(r^{3}+r_{0}^{3})^{2}$ remains strictly positive for $r > 2^{1/3}r_{0}$, guaranteeing mathematically stable orbital solutions well outside the deep quantum core. However, for radii below this critical threshold ($r \le 2^{1/3}r_{0}$), the discriminant $D$ becomes negative, rendering the roots for $\Omega_{\pm}$ complex. Physically, this mathematical breakdown signals the complete absence of bound, circular geodesics in the deep interior geometry. Any accreting matter that penetrates this boundary cannot maintain a stable Keplerian orbit and must undergo a rapid radial plunge, transitioning into a purely advection-dominated flow.

\section{Algebraic Structure of the Hayward Horizon Polynomial}
\label{app:horizons}

The causal boundaries of the regular spacetime are dictated by the roots of the modified horizon function $\Delta_{\mathcal{M}}(r) = r^{2} - 2m(r)r + a^{2} = 0$. Substituting the specific Hayward mass function from Eq.~(\ref{eq:mhay}), the horizon condition expands into a quintic polynomial:
\begin{equation}
	\label{eq:app_horizon_poly}
	r^{5} - 2Mr^{4} + a^{2}r^{3} + r_{0}^{3}r^{2} + a^{2}r_{0}^{3} = 0.
\end{equation}
To determine the analytical threshold for the existence of an event horizon, we evaluate the strictly static limit ($a=0$). The structural equation elegantly reduces to a depressed cubic polynomial:
\begin{equation}
	r^{3} - 2Mr^{2} + r_{0}^{3} = 0.
\end{equation}
The topological nature of the roots is governed by the cubic discriminant, formulated as $\Delta_{\rm cubic} = 4(r_{0}^{3})^{3} - 27(2M)^{2}(r_{0}^{3}) = 4r_{0}^{9} - 108M^{2}r_{0}^{3}$. The spacetime transitions from a black hole (possessing distinct inner and outer Cauchy horizons) to a horizonless compact object precisely at the degenerate root $\Delta_{\rm cubic}=0$. Solving this extremality condition isolates the critical static regularity parameter:
\begin{equation}
	r_{0,\rm ext} = \left(\frac{108M^{2}}{4}\right)^{\!1/6} = \frac{4M}{3\sqrt{3}} \approx 0.770\,M.
\end{equation}
For coordinate configurations exceeding this fundamental threshold ($r_{0} > r_{0,\rm ext}$), the polynomial admits no positive real roots, mathematically confirming the global regularity of the horizonless de~Sitter remnant.

\end{document}